\documentclass[pdflatex,
sn-mathphys-num,iicol]{sn-jnl}
\usepackage{graphicx}%
\usepackage{multirow}%
\usepackage{amsmath,amssymb,amsfonts}%
\usepackage{amsthm}%
\usepackage{mathrsfs}%
\usepackage[title]{appendix}%
\usepackage{textcomp}%
\usepackage{manyfoot}%
\usepackage{booktabs}%
\usepackage{algorithm}%
\usepackage{algorithmicx}%
\usepackage{algpseudocode}%
\usepackage{listings}%
\usepackage{dcolumn}
\usepackage{bm}
\usepackage[usenames,dvipsnames]{xcolor}

\begin{document}

\title[Article Title]{Single-Neutron Adding on $^{34}$S}

\author[1]{\fnm{A.~N.} \sur{Kuchera}}\email{ankuchera@davidson.edu} 
\author[2]{\fnm{C.~R.} \sur{Hoffman}}\email{calem.hoffman@gmail.com} 
\author[1]{\fnm{G.} \sur{Ryan}} 
\author[1]{\fnm{I.~B.} \sur{D'Amato}} 
\author[1]{\fnm{O.~M.} \sur{Guarinello}} 
\author[1]{\fnm{P.~S.} \sur{Kielb}} 
\author[3]{\fnm{R.} \sur{Aggarwal}} 
\author[3]{\fnm{S.} \sur{Ajayi}} 
\author[3]{\fnm{A.~L.} \sur{Conley}} 
\author[4]{\fnm{I.} \sur{Conroy}} 
\author[3]{\fnm{P.~D.} \sur{Cottle}} 
\author[3]{\fnm{J.~C.} \sur{Esparza}} 
\author[3]{\fnm{S.} \sur{Genty}} 
\author[3]{\fnm{K.} \sur{Hanselman}} 
\author[4]{\fnm{M.} \sur{Heinze}} 
\author[3]{\fnm{D.} \sur{Houlihan}}
\author[3]{\fnm{B.} \sur{Kelly}} 
\author[3]{\fnm{M.~I.} \sur{Khawaja}} 
\author[3]{\fnm{E.} \sur{Lopez-Saavedra}}
\author[3]{\fnm{G.~W.} \sur{McCann}} 
\author[3]{\fnm{A.~B.} \sur{Morelock}}
\author[4]{\fnm{L.~A.} \sur{Riley}} 
\author[3]{\fnm{A.} \sur{Sandrik}} 
\author[3]{\fnm{V.} \sur{Sitaraman}} 
\author[3]{\fnm{M.} \sur{Spieker}} 
\author[3]{\fnm{E.} \sur{Temanson}} 
\author[3]{\fnm{C.} \sur{Wibisono}} 
\author[3]{\fnm{I.} \sur{Wiedenh{\"o}ver}} 


\affil[1]{\orgdiv{Department of Physics}, \orgname{Davidson College}, \orgaddress{\city{Davidson}, \state{NC}, \postcode{28035}, \country{USA}}}
\affil[2]{\orgdiv{Physics Division}, \orgname{Argonne National Laboratory}, \orgaddress{
\city{Argonne}, \state{IL}, \postcode{60439}, \country{USA}}}
\affil[3]{\orgdiv{Department of Physics}, \orgname{Florida State University}, \orgaddress{\city{Tallahassee}, \state{FL}, \postcode{32306}, \country{USA}}}
\affil[4]{\orgdiv{Department of Physics \& Astronomy}, \orgname{Ursinus College}, \orgaddress{\city{Collegeville}, \state{PA}, \postcode{19426}, \country{USA}}}




\abstract{
\unboldmath
\textbf{Purpose:} Single-neutron adding data was collected in order to determine the distribution of the single-neutron strength of the $0f_{7/2}$, $1p_{3/2}$, $1p_{1/2}$ and $0f_{5/2}$ orbitals outside of $Z=16, N=18$, $^{34}$S.

\textbf{Methods:} The $^{34}$S($d$,$p$)$^{35}$S reaction has been measured at 8 MeV/u to investigate cross sections to excited states in $^{35}$S. Outgoing proton yields and momenta were analyzed by the Super-Enge Split-Pole Spectrograph in conjunction with the CeBrA demonstrator located at the John D. Fox Laboratory at Florida State University. Angular distributions were compared with Distorted Wave Born Approximation calculations in order to extract single-neutron spectroscopic overlaps. 

\textbf{Results:} Spectroscopic overlaps and strengths were determined for states in $^{35}$S up through 6~MeV in excitation energy. Each orbital was observed to have fragmented strength where a single level carried the majority. The single-neutron centroids of the $0f_{7/2}$, $1p_{3/2}$, $1p_{1/2}$ and $0f_{5/2}$ orbitals were determined to be $2360^{+90}_{-40}$~keV, $3280^{+80}_{-50}$~keV, $4780^{+60}_{-40}$~keV, and $\gtrsim7500$~keV, respectively.

\textbf{Conclusion:} A previous discrepancy in the literature with respect to distribution of the neutron $1p_{1/2}$ strength was resolved. The integration of the normalized spectroscopic strengths, up to 5.1~MeV in excitation energy, revealed fully-vacant occupancies for the $0f_{7/2}$, $1p_{3/2}$, and $1p_{1/2}$ orbitals, as expected. The spacing in the single-neutron energies highlighted a reduction in the traditional $N=28$ shell-gap, relative to both the $1p$ spin-orbit energy difference ($N=32$) and the lower limit on the $N=34$ shell spacing. }




\maketitle

\section{Introduction}\label{sec:intro}
The spectroscopic properties of nuclei consisting of protons occupying the $1s-0d$ shells and neutrons occupying the $1s-0d$ or $0f-1p$ orbitals are defined by intersecting behaviors. In general terms, there is a strong interplay between the spacing of the evolving (spherical) single-particle energies and the correlation energies provided by high-order, primarily quadrupole-type, excitations. The latter manifest within a single-particle framework through configurations containing various numbers of particle-hole excitations from within the $1s-0d$ shells into the $0f-1p$ shells.

An observable impact of the aforementioned competition is the Island of Inversion centered around $^{31}$Na which is characterized by a sharp transition from near-spherical ground state shapes in the $N=20$ isotones $^{36}$S and $^{34}$Si to deformed ground state shapes in $^{32}$Mg and $^{30}$Ne. The same scenario takes place along the $N=28$ isotones moving from doubly-magic $^{48}$Ca down through the deformed ground state of $^{42}$Si. Complementing the disappearance of these traditional shell closures are the appearance of new shell gaps in nuclei such as $^{24}$O ($N=16$) and $^{52,54}$Ca ($N=32,34$). The same underlying competition is connected to other phenomena in region such as stark changes in the $Z=8$ to $9$ neutron drip line locations, the presence of $p-$wave neutron-halo ground states, the postulation of (proton) bubble densities, and modifications to the spin-orbit force. Progress towards a quantitative description of the spectroscopy in the region has been discussed in Refs.~\cite{ref:Sorlin2008,ref:Heyde2011,ref:Nowaki2021,ref:Brown2022,ref:Utsuno2022} and the references therein.

Characterization of the single-neutron energy centroids of the $0f-1p$ orbitals determines their evolutionary trajectories across an isotopic chain or along isotones. The pattern or summed-strengths of the single-neutron overlap distributions provide complementary insight about the types and sizes of the correlations present throughout these systems. Recent surveys of the available single-neutron energy centroid data for the $0f-1p$ orbitals have been reported in Refs.~\cite{ref:Kay2017,ref:MacGregor2021,ref:Chen2024} across the Ne-Ca isotopes with $N=17-21$. The surveyed data leveraged the recent influx of transfer reaction data collected in inverse kinematics~\cite{ref:Brown2012,ref:Burgunder2014,ref:MacGregor2021,ref:Chen2024}, with the high-quality data available from reactions on stable-targets. In the present work, the single-neutron strength distributions and the single-neutron energy centroids were determined for the corresponding $0f-1p$ orbitals outside of $N=18$ $^{34}$S where amongst other aspects, previous experimental results have conflicted in their findings~\cite{ref:Abegg1977,ref:Piskor1984}.

We report on the single-neutron adding ($d$,$p$) reaction carried out at 8 MeV/$u$ using an enriched $^{34}$S target. The reaction populated states in $^{35}$S up to an excitation energy, $E_x$, of $\sim7.7$~MeV ($S_n = 6.986$~MeV~\cite{ref:ame2021}). The reaction probe and beam energy were ideal, in terms of the reaction kinematics and the orbital angular momentum ($\ell$) matching, for investigating overlaps with final states populated through neutron transfer into the $1s0d$-$0f1p$ shells. 
The angular distributions from measured relative differential cross sections, $d\sigma / d\Omega$, as a function of center-of-mass angle ($\theta_{cm}$), were interpreted through a Distorted Wave Born Approximation (DWBA) approach. Relative spectroscopic overlaps (spectroscopic factors), C$^2$S, were determined using known or assumed final-state total angular momenta and parities, $J^{\pi}$. Note that here, the isospin Clebsch-Gordon Coefficient, C~$=1$ where neutron-adding takes place onto an $N > Z$, isospin $T = 1$, target. Single-neutron centroid energies, $E_{cent}$, and single-neutron adding strengths, 
\begin{equation}
G_+ = \frac{2J_f+1}{2J_i+1}\textrm{C}^2\textrm{S}, 
\end{equation}
where $J_{f,i}$ are the final and initial state spins ($J_i=0$ at present), were further deduced from the $E_x$, $J^{\pi}$, $\ell$, and C$^2$S information.

\section{Background}
The current status of the observed or extracted spectroscopic properties of $^{35}$S, first discovered in 1936~\cite{ref:Thoennessen2012}, may be found in the evaluated data files of Ref.~\cite{ref:ensdf2023}. For states of interest to the present work, which lie below an excitation energy of $\sim5.1$~MeV, energies are known to better than 1~keV, and all but one state have assigned $J^{\pi}$. Previous single-neutron adding reactions on $^{34}$S fall into two classes. When $\gamma$-ray detection was used, lifetime determinations of the $^{35}$S excited levels along with $J^{\pi}$ assignments were a primary focus. A number of other dedicated $(d$,$p$) measurements were successful in determining energy levels, orbital angular momenta, and single-neutron overlap information. Of those, Refs.~\cite{ref:Moss1969,ref:Mermaz1971,ref:Abegg1977,ref:Piskor1984} are noted specifically due to discrepancies between results, for completeness in terms of excitation energy coverage, or for their closeness in reaction energy to the present work.

As the present work extracts the spectroscopic overlap between the $^{34}$S ground state and excited levels in $^{35}$S, we note that $^{34}$S has been well studied to date~\cite{ref:ensdf2023}. The ground state is known to consist of both protons and neutrons confined to the $1s-0d$ orbitals. Of particular importance to the present work is the amount of neutron $0f-1p$ shell contributions in the ground state. Analyses from data collected in single-neutron removal reactions on the $^{34}$S ground state, including both the ($p$,$d$)~\cite{ref:Moalem1975} and ($d$,$t$)~\cite{ref:Khan1988} reactions, show that less than a few percent of the total neutron removal cross section contains $\ell=1$ or $3$ contributions. Hence, the ground state contains only fractions ($\lesssim0.1$) of $0f-1p$ neutron occupancies.

\section{Experiment}\label{sec:exp}

A 16-MeV deuteron beam (8~MeV/$u$) was accelerated by the 9-MV Super-FN tandem Van-de-Graaff accelerator located at Florida State University's John D. Fox Superconducting Linear Accelerator Laboratory. Typical deuteron beam currents were on the order of 12~electrical nano-Amperes. The beam impinged on a carbon-backed sulfur target that was enriched in $^{34}$S. There was no known information associated with the target thickness or the precise enrichment of the $^{34}$S isotope. The presence of the more naturally-abundant isotope $^{32}$S was estimated to be on the order of $\approx 1 / 10$ that of $^{34}$S based on the reaction-yield information collected as discussed below. The experimental techniques and layout discussed below are analogous to those described in Refs.~\cite{ref:Riley2021,ref:Riley2022,ref:Riley2023,ref:Spieker2023,ref:Hay2024,ref:Conley2024} where additional details may be found. 

The target was positioned upstream of the Super-Enge Split-Pole Spectrograph (SE-SPS). The solid angle acceptance of the SE-SPS was fixed at 4.62~msr. Continuous beam current integration took place with a Faraday Cup located downstream of the target centered on the beam direction. The two dipole magnets of the SE-SPS provide a focusing field for the charged reaction products towards its focal plane detector suite where the protons were spatially dispersed based on their momentum. Three magnetic field settings of 7.4~kilo-Gauss (kG), 7.8~kG, and 8.4~kG were chosen to ensure overlapping and complete excitation energy coverage of $^{35}$S states through $E_x \approx 7.7$~MeV. In addition, the SE-SPS was positioned to collect data at six different laboratory scattering angles, $\theta_{lab} = 10^{\circ}, 15^{\circ}, 20^{\circ}, 30^{\circ}, 35^{\circ}$, and $50^{\circ}$, for two of the magnetic field settings (7.4~kG and 7.8~kG), and a sub-set of those angles ($\theta_{lab} = 15^{\circ}, 35^{\circ}$, and $50^{\circ}$) for the third magnetic field setting (8.4~kG).

\begin{figure}[htb]
    \centering
    \includegraphics[width=0.48\textwidth]{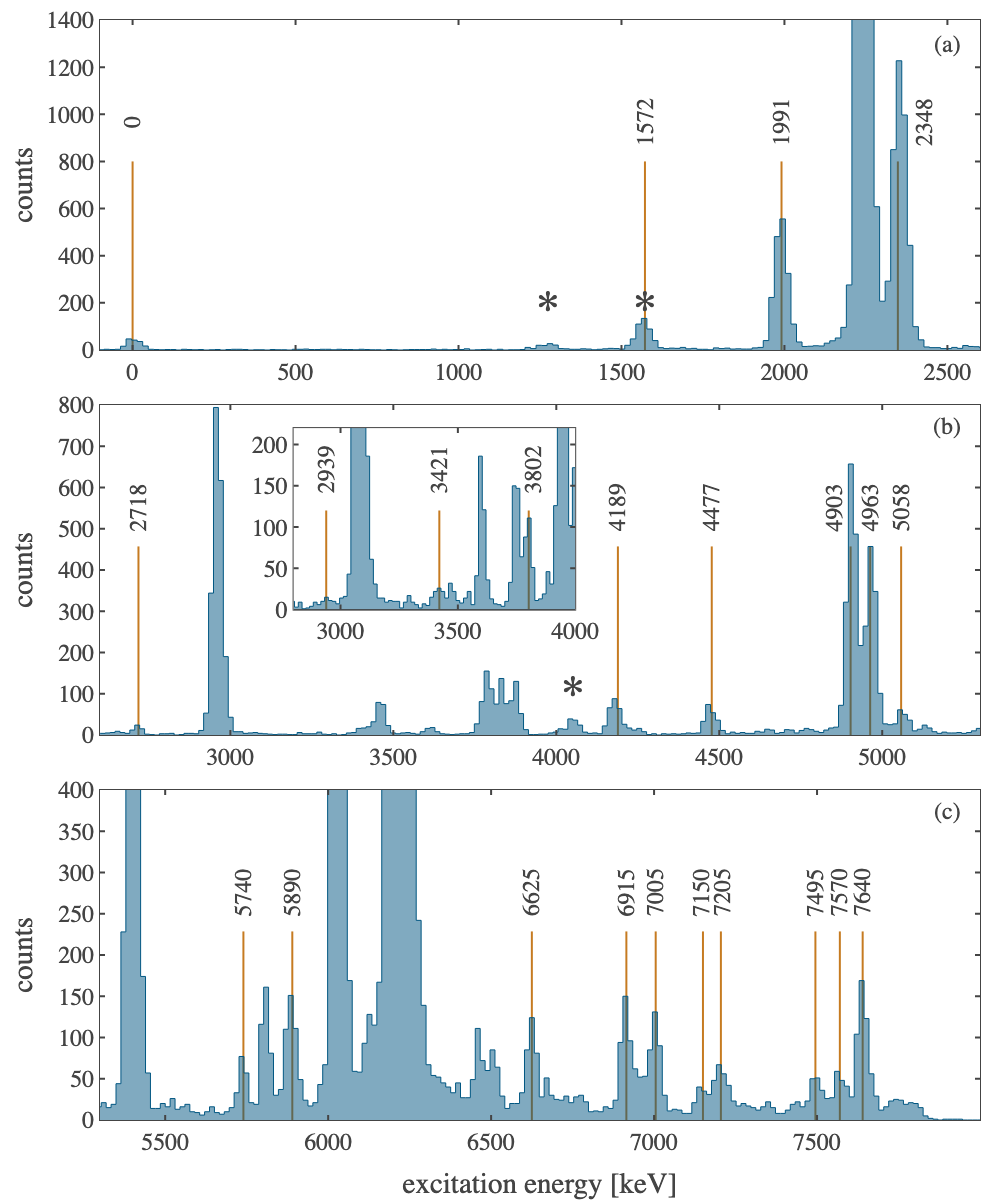}\\
    \caption{(a - c) Typical excitation energy spectra for the $^{34}$S($d$,$p$)$^{35}$S reaction at 8~MeV/$u$ and with the SE-SPS at $\theta_{lab}=20^{\circ}$. The inset spectrum within (b) is for a sub-range of excitation energy and data that were collected at $\theta_{lab}=30^{\circ}$. States identified as belonging to $^{35}$S have been labeled by their excitation energy. The locations of a few contaminant levels belonging to $^{33}$S, populated via the $^{32}$S($d$,$p$) reaction, have been labeled by an asterisk (``*") symbol. Other peaks in the yield that go unlabeled throughout (a - c) are due to reactions on C and O contaminants in the target.
    }
    \label{fig:fig1}
\end{figure}

The focal-plane detector suite consisted of a position-sensitive ionization chamber and a plastic scintillator~\cite{ref:Spieker2023}. The identification of protons from the ($d$,$p$) reactions was done by selecting the appropriate regions in the $\Delta E - E$ spectra generated from the energy loss within the ion chamber ($\Delta E$) and remaining energy detected in the scintillator ($E$). Proton-gated $Q$-value spectra were produced from the measured (dispersive) position information in the ionization chamber. Furthermore, the $Q$-value spectra were converted into calibrated $E_x$ spectra through a linear fit to the known level information on $^{35}$S up to $E_x = 5.1$~MeV~\cite{ref:ensdf2023}. Typical resolutions for the $E_x$ spectra were on the order of $\sim45$~keV FWHM. Example spectra collected for $^{35}$S are shown in Fig.~\ref{fig:fig1}.

\begin{figure}[htb]
    \centering
    \includegraphics[width=0.48\textwidth]{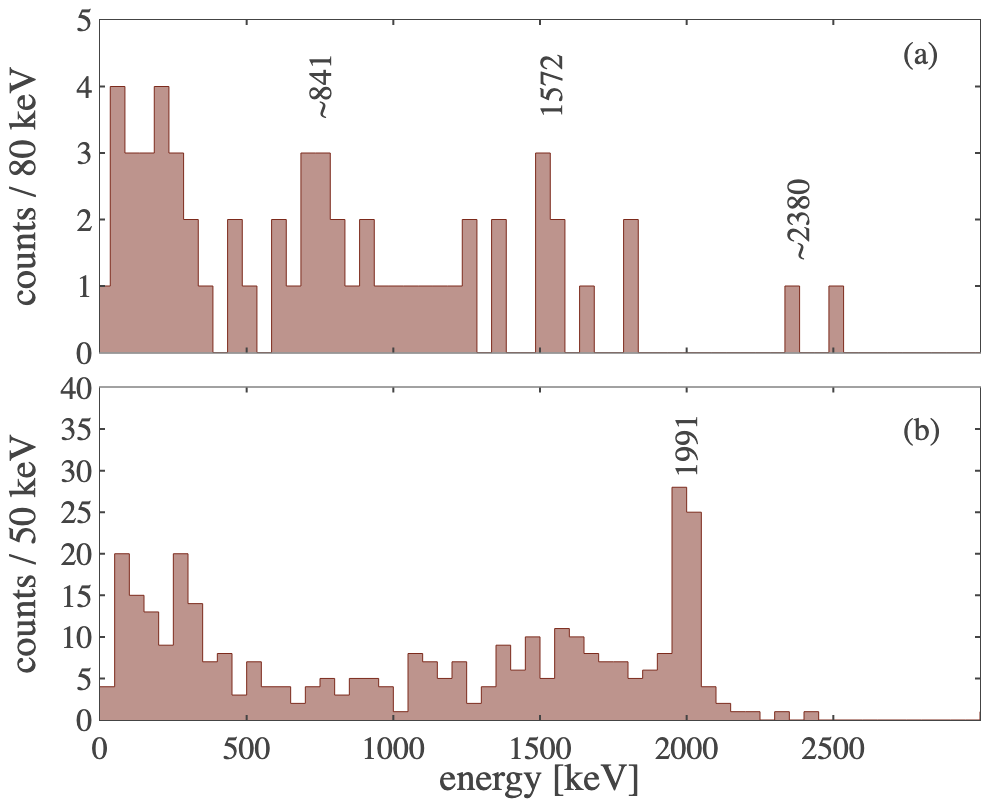}\\
    \caption{Excitation-energy gated coincident $\gamma$-ray spectra for protons corresponding to (a) $E_x = 1.571$~MeV in $^{35}$S and $E_x = 3.221$~MeV in $^{33}$S, and (b) $E_x =1.991$~MeV in $^{35}$S. The spectra comprise the integration of events across all of the detectors of the CeBrA demonstrator. Protons were accepted at $\theta_{lab} = 37^{\circ}$ by the SE-SPS.}
    \label{fig:fig2}
\end{figure}

An additional set of data for the same $^{34}$S($d$,$p$) reaction at 8~MeV/$u$ was taken with the CeBr$_3$ Array (CeBrA) demonstrator surrounding the reaction target of the SE-SPS~\cite{ref:Conley2024}. In this measurement, the outgoing protons from the $(d$,$p)$ reaction were detected in coincidence with de-exciting $\gamma$ rays (Fig.~\ref{fig:fig2}). These data were primarily used in the present work to provide information on some overlapping peaks produced from the $^{32}$S contaminants in the target. The CeBrA demonstrator was comprised of two sizes of CeBr$_3$ detectors, $2''\times2''$ and $3''\times4''$. The SE-SPS was set to a fixed angle of $\theta_{lab}=37^{\circ}$ and had a magnetic field setting of 7.8~kG. Figures~\ref{fig:fig2}(a) and (b) display the coincident $\gamma$-ray energy spectra with selected excitation energies in $^{35}$S. In particular, Fig.~\ref{fig:fig2}(b) highlights the known 7/2$^-\rightarrow3/2^+$ 1991-keV transition in $^{35}$S which has essentially a 100\% branch from the 1.991-MeV state to the ground state~\cite{ref:Grocutt2022}. No additional $\gamma$-ray transitions are present in the spectrum as this state is well isolated from any anticipated contaminants in the SE-SPS focal plane. A confirmation of the lifetime of the 1.991-MeV level was also made from the present data as reported in Ref.~\cite{ref:Conley2024}.

\begin{table}[!htb]
\caption{The single-neutron spectroscopic overlaps (C$^2$S) and strengths ($G_+$) from the $^{34}$S($d$,$p$)$^{35}$S reaction at 8~MeV/$u$. Uncertainties from both the fitting results as well as sensitivities to the DWBA optical model parameters ($5-10$\% for $\ell>0$ and $10-15$\% for $\ell=0$) were included. The C$^2$S have been normalized through summed strengths as described in the text. The excitation energies ($E_x$) listed without uncertainties or without parenthesis on $J^{\pi}$, were adopted from previous work~\cite{ref:ensdf2023}. Only the upper limits on the strengths have been listed for states above $E_x = 6$~MeV due to ambiguities in the DWBA fit results and unknown $J^{\pi}$ values.\label{tab:tab1}
}
\begin{tabular}{@{}lccll}
\toprule
$E_x$ [MeV] & $J_f$ & $\ell$ & C$^2$S & $G_{+}$ \\
\midrule
0 & 3/2$^+$ & 2  & 0.54(12) & 2.2(5)\\
1.571 & 1/2$^+$ & 0  & 0.14(7)\footnotemark[1]   & 0.3(2)\\
1.991 & 7/2$^-$ & 3  & 0.87(9) & 7.0(7)\\
2.348 & 3/2$^-$ & 1  & 0.62(7) & 2.5(3) \\
2.718 & 5/2$^+$ & 2  & 0.02(1) & 0.1(1)\\
2.939 & 3/2$^+$ & 2   & 0.01(1)\footnotemark[2] & $<0.1$\\
3.421 & 5/2$^+$ & 2   & 0.02(1)\footnotemark[2] & 0.1(1)\\
3.802 & 3/2$^-$ & 1   & 0.07(4)\footnotemark[2] & 0.3(2)\\
4.189 & 1/2$^-$ & 1  & 0.17(2) &  0.3(1)\\
\multirow{2}{*}{4.477\footnotemark[3]} & 3/2$^{+}$ & 2  & 0.07(2) & 0.3(1)\\
& 7/2$^-$ & 3  & 0.05(2) & 0.4(2)\\
4.903 & 1/2$^-$ & 1  & 0.78(8) & 1.6(2)\\
4.963 & 3/2$^-$ & 1  & 0.32(3) & 1.3(1)\\
5.058 & 7/2$^-$ & 3  & 0.08(1) & 0.6(1)\\
5.740(20) & (5/2$^-$) & 3  & 0.03(1) & 0.2(1)\\
5.890(20) & (3/2$^+$) & 2  & 0.11(2) & 0.4(1)\\
\midrule
\multirow{2}{*}{6.625(30)} & \multirow{2}{*}{--} & (2)  & \multirow{2}{*}{--} & $<0.3$\\
&  & (3)  &  & $<0.4$\\
\multirow{2}{*}{6.915(30)} & \multirow{2}{*}{--} & (1)  & \multirow{2}{*}{--} & $<0.3$\\
&  & (3)  &  & $<0.4$\\
\multirow{2}{*}{7.005(35)} & \multirow{2}{*}{--} & (2)  & \multirow{2}{*}{--} & $<0.3$\\
&  & (3)  &  & $<0.4$\\
\multirow{2}{*}{7.150(35)} & \multirow{2}{*}{--} & (0)  & \multirow{2}{*}{--} & $<0.3$\\
&  & (3)  &  & $<0.1$\\
\multirow{2}{*}{7.205(35)} & \multirow{2}{*}{--} & (1)  & \multirow{2}{*}{--} & $<0.1$\\
&  & (2)  &  & $<0.2$\\
\multirow{2}{*}{7.495(40)} & \multirow{2}{*}{--} & (1)  & \multirow{2}{*}{--} & $<0.1$\\
&  & (2)  &  & $<0.1$\\
\multirow{2}{*}{7.570(40)} & \multirow{2}{*}{--} & (2)  & \multirow{2}{*}{--} & $<0.1$\\
&  & (3)  &  & $<0.1$\\
\multirow{2}{*}{7.640(40)} & \multirow{2}{*}{--} & (2)  & \multirow{2}{*}{--} & $<0.3$\\
&  & (3)  &  & $<0.4$\\
\bottomrule
\end{tabular}
\footnotetext[1]{Extracted from a multi-distribution fit to the data to account of the $\ell=1$ transfer to the $^{33}$S 3.22~MeV state.}
\footnotetext[2]{Based on measured yields from only one or two SE-SPS angle settings.}
\footnotetext[3]{Assumed to be a doublet within the $E_x$ resolution at this energy.}
\end{table}

\section{Methods and Results}\label{sec:methres}
\subsection{Levels in $^{35}$S}

The states in $^{35}$S residing below 5.1~MeV in excitation energy, known from previous ($d$,$p$) measurements, have been identified in the calibrated $E_x$ spectra of Fig.~\ref{fig:fig1} and Table~\ref{tab:tab1}. No additional levels were observed up through this $E_x$. Previously unknown states have been noted above $E_x > 5.1$~MeV in the figure and the table. The larger uncertainties on these $E_x$ are due to the extrapolation of the linear energy calibration. Of these additional states, a few have overlapping energies with known levels~\cite{ref:ensdf2023}, for instance from past $(d$,$p$)~\cite{ref:Abegg1977}, neutron capture~\cite{ref:Carlton1984}, or ($p$,$^{3}$He)~\cite{ref:Guichard1975} measurements. However, no precise correspondences were concluded. For some levels, yields may have been discarded for a specific SE-SPS angle setting or two due to an overlapping contaminant line. Such contaminants in the excitation energy spectra were identified through their calculated kinematics and expected placement within the spectra, or by their energy dependence as a function of SE-SPS angle setting. In a complementary fashion, new levels suggested to belong to $^{35}$S were considered only if they appeared at the same $E_x$ ($\sim$few keV) for at least three SE-SPS angle settings.

The only state found to be directly impacted by a $^{33}$S contaminant level was the $E_x = 1.572$-MeV state. As was the case in previous work (for example see Fig. 1 of Ref.~\cite{ref:Moss1969}), it was similarly determined to be a doublet including the 3/2$^-$ ($\ell=1$) state at 3.221 MeV in $^{33}$S. This determination was made both from the kinematic calculations of the ($d$,$p$) reaction and the supporting information provided by the $(d$,$p\gamma$) data. As viewed in Fig.~\ref{fig:fig2}(a), an expected 1/2$^+\rightarrow3/2^+$ 1572-keV transition is observed in the $E_x = 1.572$-MeV gated $\gamma$-ray spectrum. However, in addition, an $\sim840$-keV transition is also apparent. This transition belongs to the 3/2$^-\rightarrow1/2^+\rightarrow3/2^+$, $2380\rightarrow841$-keV, cascade in $^{33}$S. Due to low statistics and $\gamma$-ray detection efficiencies at higher energies, the 2380-keV transition, as well as the ground state 3221-keV transition, were not observed.

Therefore, as discussed in more detail below, a multi-distribution fit was applied to the angular distribution data for the 1.572-MeV state. Other states in $^{33}$S, which did not interfere with $^{35}$S levels, were also confirmed in the $E_x$ spectra and their C$^2$S values have been extracted.

\subsection{Angular distributions}

The yields for identified states were collected at each SE-SPS angle setting when no contaminant line(s) interfered, with the exception of the $E_x = 1.572$~MeV state. The differential cross sections, $d\sigma / d\Omega$, were determined by normalizing to the integrated charge collected over the duration of the data-collection period. The constant value of the SE-SPS solid angle acceptance was applied, though not strictly needed, as an absolute scale on the $d\sigma / d\Omega$ was not possible due to the unknown amount of $^{34}$S isotope content in the target. When levels of interest overlapped across multiple SE-SPS field settings, the $d\sigma / d\Omega$ from each setting were combined in an un-weighted average. Each $\theta_{lab}$ scattering angle setting of the SE-SPS was converted into $\theta_{cm}$ through the known kinematics of the reaction. 

\begin{figure}[htb]
    \centering
    \includegraphics[width=0.48\textwidth]{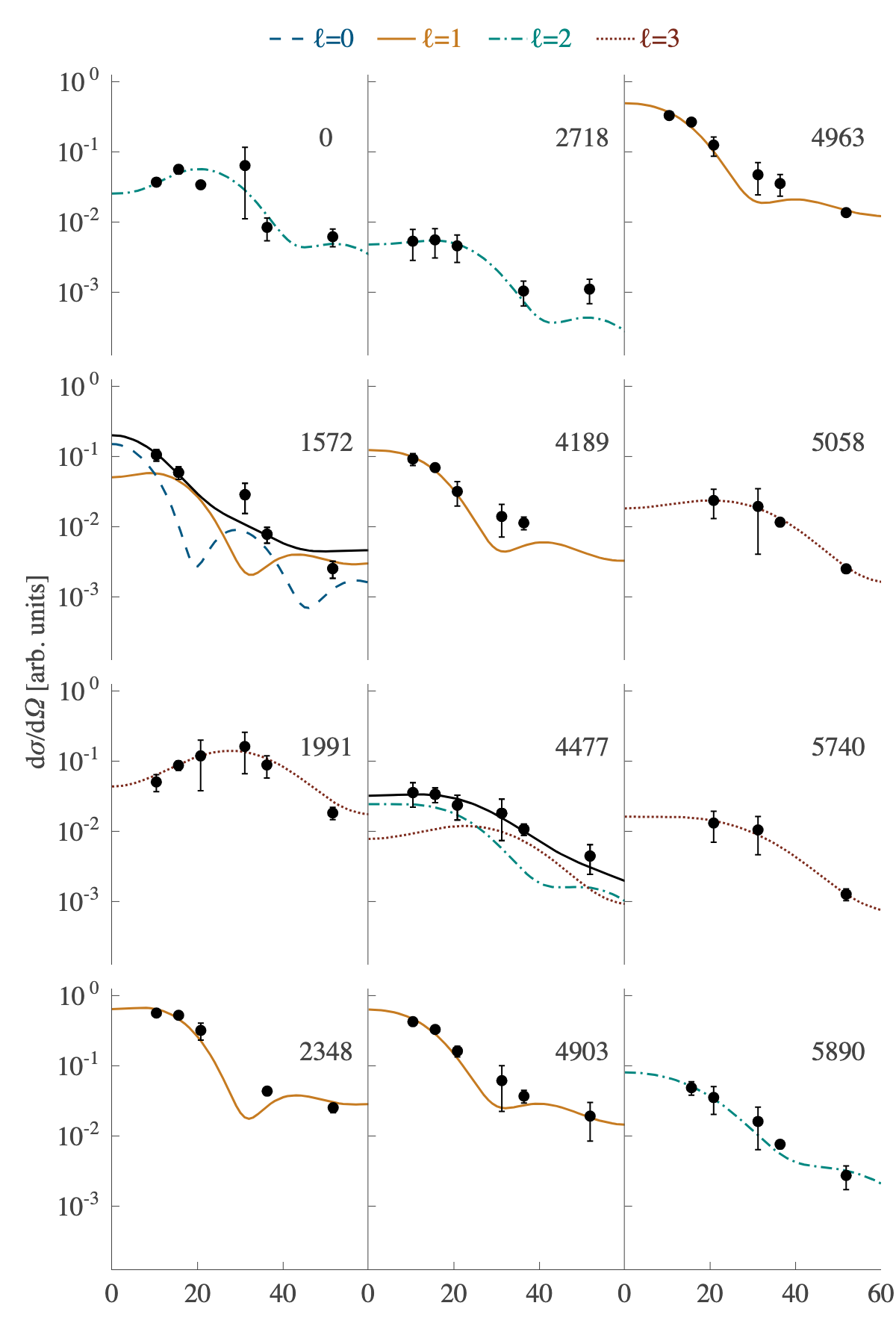}\\
    \caption{The $^{34}$S($d$,$p$)$^{35}$S differential cross sections, d$\sigma$/d$\Omega$ [arb. units] are given as a function of center-of-mass angle, $\theta_{cm}$, by the black data points. Each angular distribution has been labeled by its excitation energy in keV. Uncertainties on the data are represented by the error bars when they are larger than the size of the symbol and they include both the statistical and 15\% systematic contributions (see text for additional details). The calculated distributions based on the DWBA prescription described in the text are represented by the lines and include four possible $\ell$ transfer values (0 - dark-blue, dashed, 1 - yellow, solid, 2 - blue-green, dot-dashed, and 3 - red, dotted). The resulting C$^2$S from the fits of the calculated to the experimental angular distributions are listed in Table~\ref{tab:tab1}. }
    \label{fig:fig3a}
\end{figure}

\begin{figure}[htb]
    \centering
    \includegraphics[width=0.32\textwidth]{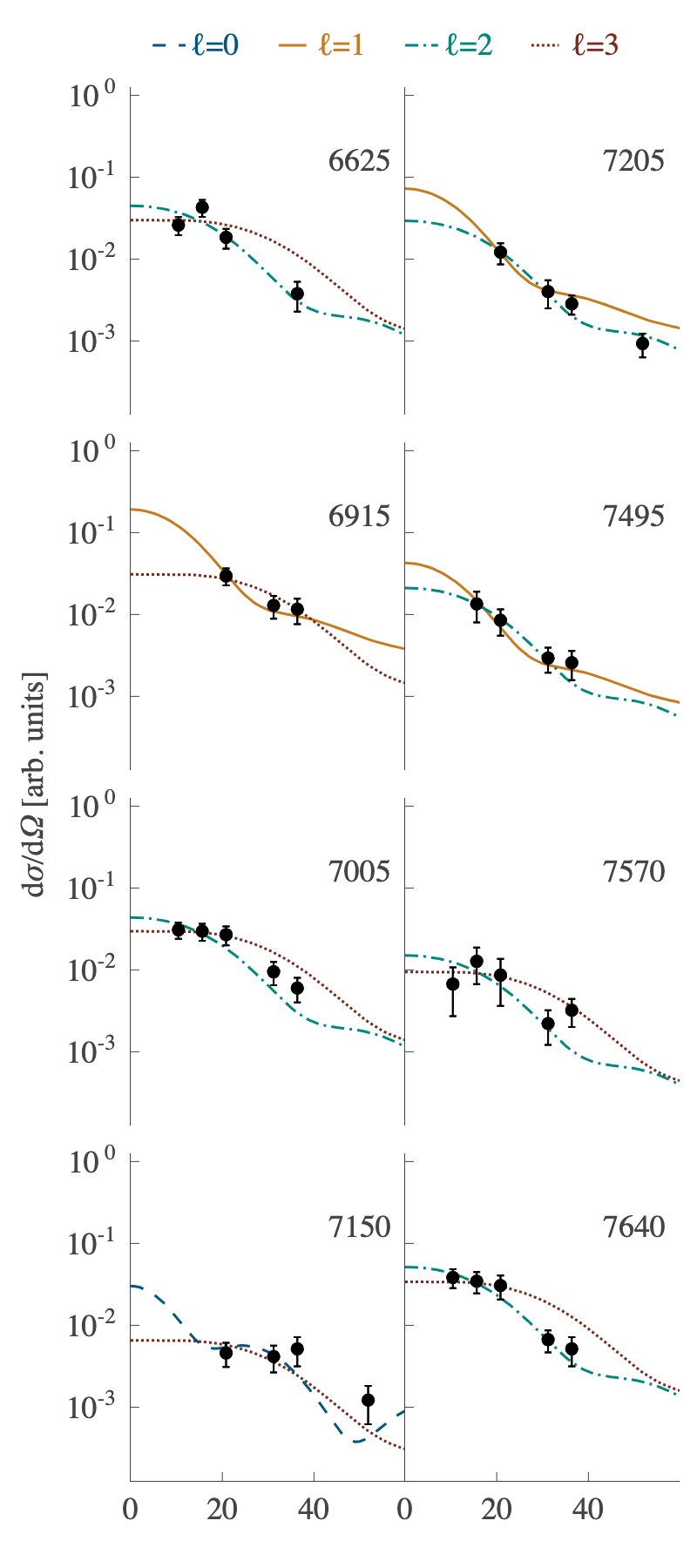}\\
    \caption{Same as Fig.~\ref{fig:fig3a} for additional levels above $E_x = 6$~MeV.}
    \label{fig:fig3b}
\end{figure}

Experimental angular distributions are shown in Figs.~\ref{fig:fig3a} and ~\ref{fig:fig3b} for $^{35}$S levels in which yields could be measured in at least three SE-SPS angles. The horizontal error bars on the data are smaller than the points size. The vertical error bars on the $d\sigma / d\Omega$ values consist of uncertainties associated with the yield determinations as well as a previously characterized systematic uncertainty of 15\% due to the charge integration~\cite{ref:Riley2021,ref:Riley2022,ref:Riley2023,ref:Spieker2023,ref:Hay2024}. The statistical contribution was only sizeable for the weaker states or those above $E_x \approx 6$~MeV. Uncertainties due to non-uniformity in the target thickness and changes in the deuteron beam spot parameters are expected to be small and contained within the 15\% uncertainty already applied. 

\subsection{Distorted Wave Analysis}

Theoretical angular distributions representing the single-neutron transfer cross sections as a function of $\theta_{cm}$ were calculated using the Distorted Wave Born Approximation (DWBA) approach. The calculations were done with the finite-range DWBA software package \textsc{PTOLEMY}~\cite{ref:Macfarlane78}. The Argonne $v_{18}$ potential~\cite{ref:Wiringa1995} was included to represent the deuteron bound-state wave function. A Woods-Saxon potential was used for the final bound-state wave function of the neutron with potential parameters of $R=r_0\textrm{A}^{1/3}$ [$r_0 = 1.25$~fm], $a = 0.65$~fm, and $R^{C}=r^{C}_0\textrm{A}^{1/3}$ [$r^{C}_0 = 1.3$~fm], as well as spin-orbit parameters of $V_{so} = 6.0$~MeV, $R^{so}=r^{so}_0\textrm{A}^{1/3}$ [$r^{so}_0 = 1.1$~fm], and $a_{so} = 0.65$~fm. The potential depth was varied to reproduce the binding energies of the states in $^{35}$S below $E_x=6.5$~MeV. Above this energy, where only upper limits on the C$^2$S values were determined, a fixed binding energy corresponding to $E_x = 6.5$~MeV was used. This was due to the asymptotic behaviour of low-$\ell$ orbitals in proximity to the one-neutron separation energy ($S_n = 6.986$~MeV~\cite{ref:ame2021}). The choice of bound-state radius affected the \textit{relative} spectroscopic factors by less than 5\%. The distorting optical potentials of the incoming deuteron and outgoing proton were parameterized through the global potentials defined in Refs.~\cite{ref:An2006} and ~\cite{ref:Koning2003}, respectively. Other global optical potential parametrizations, such as those defined in Refs.~\cite{ref:Perey1963d,ref:Perey1963p,ref:Perey1976}, produced relative C$^2$S values within the quoted uncertainties listed in Table~\ref{tab:tab1}.

The resulting fits of the calculated angular distributions to the measured angular distributions are shown in Figs.~\ref{fig:fig3a} and ~\ref{fig:fig3b}. For the negative-parity $^{35}$S states below $E_x < 5.1$~MeV only the 4.477~MeV state(s) did not have a known $J^{\pi}$~\cite{ref:ensdf2023}. Each fit was consistent with the expected $\ell$ transfer value for the known $J^{\pi}$ (Fig.~\ref{fig:fig3a}). A multi-distribution fit was used for the 4.477-MeV state due to conflicting information in the literature and the possibility for a level doublet. A previous ($d$,$p$) measurement suggested $J^{\pi}=7/2^-$ based on the angular distribution~\cite{ref:Abegg1977}. In contrast, $(d$,$p\gamma$) data has shown multiple $\gamma$-ray branches to low-lying states including the 1/2$^+$ ground state~\cite{ref:VanderMark1972,ref:Freeman1972} which would not be favored from a 7/2$^-$ level, and instead, suggests a $J^{\pi} \leq 5/2^+$ for the level. A combined $\ell=2$ plus $\ell=3$ multi-distribution fit reproduced the data and therefore a doublet has been adopted throughout this work.

For $^{35}$S states above $E_x > 5.1$~MeV which could not be associated with any previously known levels or known $J^{\pi}$, angular distributions considering $0 \leq \ell \leq 3$ neutron transfers were independently fit to the angular distributions. No multi-$\ell$ fit combinations were considered. Only the 5.740~MeV and 5.890~MeV levels produced distinct results (Fig.~\ref{fig:fig3a}). For states above $E_x > 6$~MeV, the fits which produced reasonable $\chi^2$ values have been displayed in Fig.~\ref{fig:fig3b}. 

In the case of the 1.571-MeV state in $^{35}$S, due to the discussed overlap with the $^{33}$S 3.221-MeV state populated from the $\ell=1$ ($1p_{3/2}$) neutron transfer on the $^{32}$S in the target, a multi-parameter fit was performed on the measured angular distribution. The DWBA angular distributions for the $^{33}$S $\ell=1$ component was calculated for the proper kinematics of the $^{32}$S($d$,$p$)$^{33}$S reaction at the same beam energy.

\section{Discussion}

The $^{34}$S($d$,$p$)$^{35}$S spectroscopic overlaps, C$^2$S, determined from fits of the DWBA angular distributions to the experimental angular distributions (Figs.~\ref{fig:fig3a} and ~\ref{fig:fig3b}), have been listed in Table~\ref{tab:tab1} for the states having established $\ell$ values ($E_x \lesssim 6$~MeV). In the cases of the 5.740(20)~MeV and 5.890(20)~MeV levels, the $J^{\pi}$ assignments are purely speculative. The uncertainties on the C$^2$S values incorporate variations in the adopted DWBA parameters and were on the order of 5\%-10\% for $\ell > 0$ transfer and $\sim$10\%-15\% for any $\ell=0$ distributions. Further uncertainties from the choice and outcome of the angular distribution fits were included and were on the order of $\sim5 - 10$\% for the key states of interest below $E_x < 5.1$~MeV (excluding the multi-distribution fits).

For the $\ell=1$, $3/2^-$ state at $E_x = 3.802$~MeV, only the yield from a single angle was measured. In this case, the C$^2$S was extracted by first normalizing the 3.802~MeV yield to that of the $\ell=1$ 4.963-MeV level at the same angle. An additional correction was then applied based on the ratio of the DWBA cross sections for the two different $E_x$ values at the corresponding angle. The uncertainty for the C$^2$S value was dominated by the statistical uncertainty of the 4.963-MeV data point. The same procedure was applied for the more weakly-populated $\ell=2$, 5/2$^+$ levels at 2.939 and 3.421~MeV, where yields for only two angles were measured. In those cases, the 2.718-MeV state provided the normalization and the final value was the average of the two points, though large uncertainties persisted due to statistics.

The C$^2$S values listed in Table~\ref{tab:tab1} have been normalized through a procedure based on the well-known Macfarlane and French sum rules~\cite{ref:Macfarlane78}. In the present case, the normalization value was determined from the $\ell=3$ strengths summed up through $E_x = 5.1$~MeV and divided by the expected $0f_{7/2}$ vacancy, i.e. $\sum G_+(\ell=3) / (2j+1)$, where $j=7/2$. The measured strength over this excitation energy range was assumed to belong solely to the $\nu0f_{7/2}$ orbital (and not the higher-lying $\nu0f_{5/2}$ orbital). The $\ell=3$ strength from the multi-distribution fit done on the 4.477~MeV-state was included, even though it only contributed $\sim$5\% the total strength. The assumption made of a fully unoccupied $0f_{7/2}$ orbital, hence, not requiring the single-neutron removal strength ($G_-$), was supported by the less than a percent relative C$^2$S values obtained through various neutron-removal reactions on $^{34}$S~\cite{ref:Moalem1975,ref:Khan1988}. The normalization value determined from the $\ell=3$ strength was commonly applied to all of the C$^2$S values, independent of $\ell$. It can also be noted, that for this specific reaction, the procedure outlined above is equivalent to normalizing the sum of the corresponding $\ell=3$, C$^2$S values to equal 1.

As noted above, states in $^{33}$S were populated in the $^{32}$S($d$,$p$) reaction due to the presence of $^{32}$S in the target. C$^2$S values were extracted for three of these states as checks of the analysis procedures as well as to estimate of the amount of relative $^{34}$S to $^{32}$S present in the target. The three levels, including the 7/2$^-$ 2.935-MeV ($\ell=3$) level, the 3/2$^-$ 3.221-MeV ($\ell=1$) level, and the 1/2$^-$ 5.711-MeV ($\ell=1$) level, appeared at $E_x \approx 1.29$~MeV, 1.57~MeV, and 4.06~MeV in the $^{35}$S $E_x$ spectra, respectively (denoted by asterisks in Fig.~\ref{fig:fig1}). The relative C$^2$S values for the 2.935-MeV 7/2$^-$ state [C$^2$S = 0.08(1)], the 3.221-MeV 3/2$^-$ state [C$^2$S = 0.06(2)], and the 5.711-MeV 1/2$^-$ state [C$^2$S = 0.07(1)] are similar to their reported values of $\sim0.5$, $\sim0.9$, and $\sim0.5$, respectively~\cite{ref:ensdf2023}. 

The ratio of the extracted $^{33}$S C$^2$S values to corresponding fragments of similar strength in $^{35}$S, i.e. those with C$^2$S~$\gtrsim$~0.6, sets the contributions of the $^{33}$S states at $<10$\% of those of $^{35}$S. This sets the $^{34}$S/$^{32}$S target material ratio on the order of $\approx10$. It also confirms that non-identified strength from any other $^{33}$S levels would contribute negligibly to the $^{35}$S yields.

\begin{figure}[htb]
    \centering
    \includegraphics[width=0.48\textwidth]{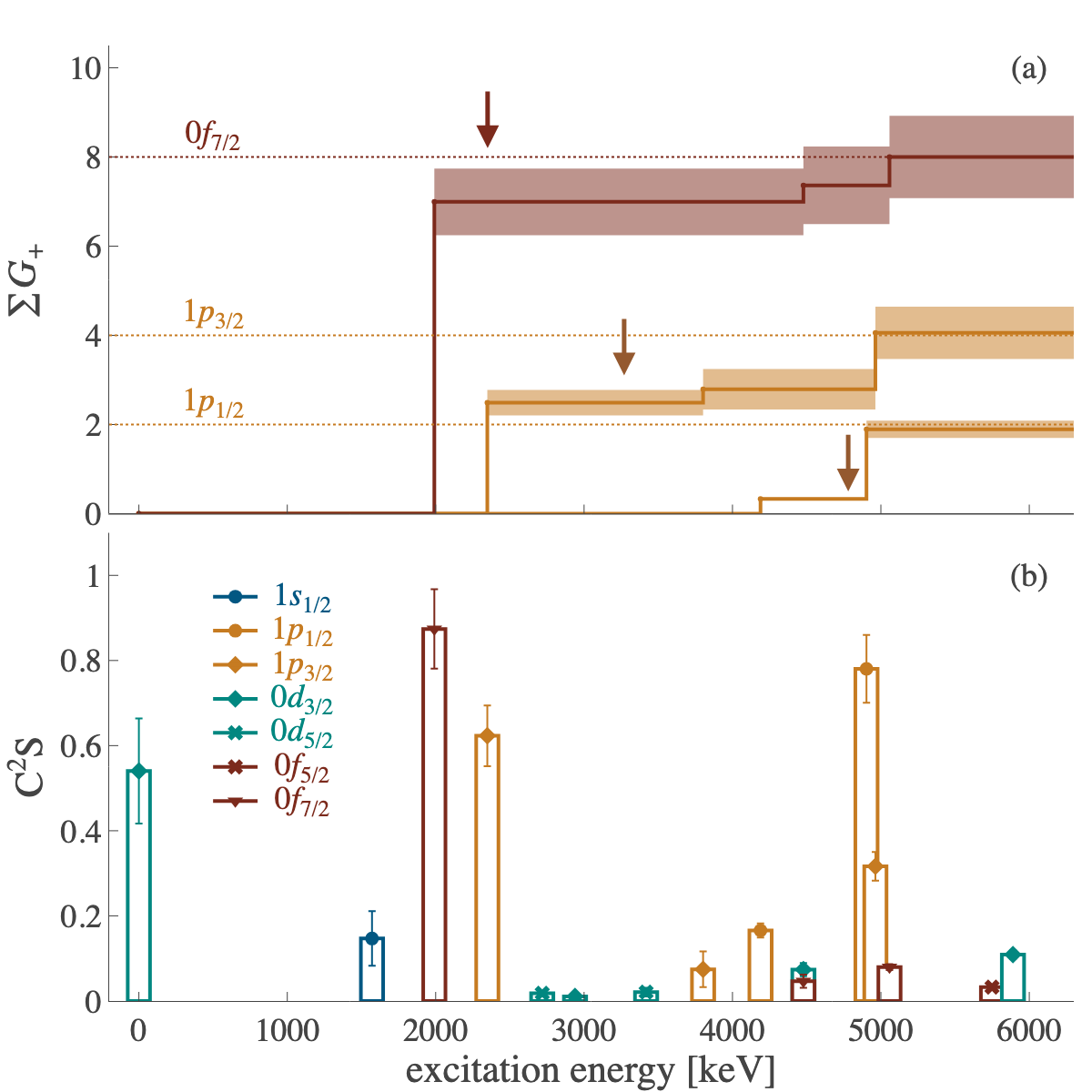}\\
    \caption{(a) Integrated experimental single-neutron strength ($\sum G_{+}$) as a function of excitation energy ($E_x$) in $^{35}$S. The horizontal lines indicate the total vacancy expected for each orbital. Note that the integrated $0f_{7/2}$, $\ell = 3$ strength was used to determine the normalization factor of the C$^2$S, i.e. $\sum G_+ = 8$ was ensured for this orbital. Arrows indicate the single-neutron energy centroid energies for each corresponding orbital. (b) The distribution of the extracted single-neutron spectroscopic strength (C$^2$S) in $^{35}$S up through $E_x \approx 6$~MeV (Table~\ref{tab:tab1}). Uncertainties are shown when larger than the data points and each orbital is distinguished by color and marker type.}
    \label{fig:fig4}
\end{figure}
The validity of a commonly applied normalization was demonstrated through an inspection of the normalized integrated strength for the neutron $1p_{3/2}$ and $1p_{1/2}$ orbitals. Over the same excitation energy range, $E_x \leq 5.1$~MeV, $\sum G_+$ produced values of 4.1(6) and 1.9(3), for the $1p_{3/2}$ and $1p_{3/2}$ orbitals, respectively. These are in agreement with the vacancy expectations of $4$ and $2$. The evolution of these summed strengths as a function of $E_x$ are presented in Fig.~\ref{fig:fig4}(a). Congruently, the normalized summed C$^2$S values for each the $0f_{7/2}$, $1p_{3/2}$ and $1p_{1/2}$ neutron orbitals are $=1.00(12)$, 1.01(14), and 0.95(10), respectively.

While the choice of an upper limit of summed strength at $E_x = 5.1$~MeV is somewhat arbitrary, it is clear from the integrated strengths displayed in Fig.~\ref{fig:fig4}(a) that similar amounts of strengths are being equally missed across the three orbitals. Conservatively, an overall uncertainty of 10-20\% should be considered on the relative vacancies for variations in the missed strength. If all possible strength from the levels between 5.1 - 8 MeV were considered, an additional amount of normalized strength of $\lesssim 2$ may be present. This is about 15\% of the total vacancy ($8+4+2=14$) without inclusion of the $0f_{5/2}$, and about 10\% of a full vacancy of $20$ if this upper orbital is included. This additional uncertainty has not been explicitly applied to the errors in Table~\ref{tab:tab1} or Fig.~\ref{fig:fig4}(a).

The distribution of the normalized C$^2$S values is shown as a function of $E_x$ in Fig.~\ref{fig:fig4}(b). There is a fragment carrying the majority of the strength for each orbital with values of $\sim$90\%, $\sim$60\%, and $\sim$80\%, for the $0f_{7/2}$, $1p_{3/2}$ and $1p_{1/2}$ orbitals, respectively. Other fragments, ranging from $\sim$5 - 30\% of the integrated strength comprise the remaining strength below $E_x = 5.1$~MeV. A similarly dominant fragment corresponding to strength in the $\nu0f_{5/2}$ orbital was not observed through our $E_x\sim$~7.7~MeV sensitivity range.

Due to discrepancies between previous ($d$,$p$) measurements, it is worth pointing out that the present results are in agreement with the results of Ref.~\cite{ref:Abegg1977}, specifically those of potential set 2 listed in their Table 2. The present work shows disagreement with Ref.~\cite{ref:Piskor1984} where a larger fragment of strength was extracted for the $\ell=1$, $E_x = 4.963$-MeV state. It is possible that due to the large number of isotopes in their target, the additional strength did not belong to the $^{34}$S($d$,$p$) reaction.

Single-neutron centroid energies, $E_{cent}$, for the $0f_{7/2}$ ($E_{cent} = 2360^{+90}_{-40}$~keV), $1p_{3/2}$ ($E_{cent} = 3280^{+80}_{-50}$~keV), and $1p_{1/2}$ ($E_{cent} = 4780^{+60}_{-40}$~keV) orbitals were determined from their ($2J+1$)C$^2$S-weighted excitation energy distributions. The $E_{cent}$ values are indicated in Fig.~\ref{fig:fig4}(a) by arrows for each corresponding orbital. Only the levels below $E_x < 5.1$~MeV determined the $E_{cent}$, however, their associated errors reflect possible contributions from higher-lying strength (Table~\ref{tab:tab1}). The lack of sizeable $\ell=3$ strength which could be associated with the neutron $0f_{5/2}$ orbital, places a lower limit on its $E_{cent}$ at $\gtrsim 7.5$~MeV.

The $E_{cent}$ are not designated as the effective single-particle energies considering the energy-weighted summation did not include the single-neutron removal strengths~\cite{ref:Macfarlane78}, albeit they have been found to be small~\cite{ref:Moalem1975,ref:Khan1988}. It should be emphasized, that due to the observed fragmentation of the single-neutron strength, if one were to incorrectly interpret \emph{only} the major fragments as the centroid energies for the $0f_{7/2}$, $1p_{3/2}$ and $1p_{1/2}$ orbitals, the inappropriate $E_{cent}$ energies would be different by $\sim350$~keV, $\sim900$~keV, and $\sim120$~keV, respectively.

The energy spacing between the orbitals is an indication of the spherical neutron shell gaps outside of $N=18$, $^{34}$S. The neutron $1p_{3/2} - 0f_{7/2}$ centroid energy difference is 920$^{+120}_{-140}$~keV and is the spacing for the traditional $N=28$ shell gap. Similarly, the $1p_{1/2} - 1p_{3/2}  = 1500^{+110}_{-120}$~keV energy difference reflects the neutron $1p$ spin-orbit partner splitting. Its size is also that of the $N=32$ sub-shell gap, assuming a normal ordering and filling of the $1p_{3/2}$ orbital. Relative to the upper most $\nu1p_{1/2}$ centroid, the estimated $0f_{5/2} - 1p_{1/2}$ energy spacing reflecting the $N=34$ gap is $\gtrsim2.5$~MeV. This lower limit is larger than both the traditional $N=28$ gap and the ($N=32$) spin-orbit partner energy difference in this system. For completeness, we note that the limit on $0f$ spin-orbit energy spacing is $0f_{5/2} - 0f_{7/2} \gtrsim 5000$~keV. 

To make a comparison, we explore the corresponding $N=21$ single-neutron energy gaps outside of $N=20$, $^{36}$S and based on the C$^2$S data extracted in Ref.~\cite{ref:Eckle1989}. They return energy separations of $\approx1200$~keV for $N=28$, $\approx1500$~keV for $N=32$, and $\gtrsim2700$~keV for $N=34$, similar to those extracted in the present work for orbitals outside of $^{34}$S. Although $^{34,36}$S are stable isotopes they each have a reduced $N=28$ traditional shell-gap spacing relative to their corresponding $N=32$ and $34$ gaps. A reduced $N=28$ energy gap is known to persist in the $Z=16$, $N\approx28$ neutron-rich isotopes being one of the key components leading to the lower-lying intruder configurations in $^{44}$S (and $^{42}$Si)~\cite{ref:Sorlin2008,ref:Heyde2011,ref:Nowaki2021,ref:Brown2022,ref:Utsuno2022}.

The energy ordering of the orbitals outside of $^{34}$S and $^{36}$S follows those found in the Ca isotopes. In both $^{41}$Ca and $^{49}$C~\cite{ref:ensdf2023}, the neutron $0f_{5/2}$ centroid resides at sizeable energy ($>1$~MeV) above the neutron $1p_{1/2}$ energy, thus generating an $N=34$ sub-shell gap. Similarly, data collected on a number of neutron-rich $N=17, 19$ and $21$ isotopes, including those in the present work, demonstrate the persistence of this sub-shell spacing down through at least $Z=14$~\cite{ref:Burgunder2014,ref:Chen2024}. The $1p_{1/2} \rightarrow 0f_{5/2}$ ordering is in opposition to the level-energy sequence determined for the lowest $1/2^-$ and $5/2^-$ states in $N\approx Z$ $^{57}$Ni~\cite{ref:Rehm1998}, which implies reversed or near-degenerate $0f_{5/2}-1p_{1/2}$ neutron orbital energies (Fig.~5 of Ref.~\cite{ref:Otsuka2020}). Recent work has also revisited the available data which determines the evolution of these orbitals as a function of proton $0f_{7/2}$ occupancy ($Z = 20 - 28$) in the $N=29$ isotones~\cite{ref:Riley2021,ref:Riley2022,ref:Riley2023}. Further discussion points on the underlying mechanisms which produce the energy shifts are contained within Refs.~\cite{ref:Sorlin2008,ref:Heyde2011,ref:Nowaki2021,ref:Brown2022,ref:Utsuno2022}.


Considering the single-neutron centroid energies relative to the $^{35}$S one-neutron separation energy, $S_n - E_{cent}$, where $S_n = 6.986$~MeV~\cite{ref:ame2021}, provides some perspective on influences from weak-confinement or continuum-couplings on their relative energies. The $S_n - E_{cent}$ values for the three orbitals of interest are roughly, $4630$~keV ($0f_{7/2}$), $3710$~keV ($1p_{3/2}$), $2210$~keV ($1p_{1/2}$), and $\lesssim 0$~keV ($0f_{5/2}$). The two $1p$ orbitals have the additional confinement of $\sim$1~MeV beyond the particle threshold due to their $\ell=1$ orbital angular momentum barriers. The $\ell=3$ orbitals even more so, with barriers of $\gtrsim 3$~MeV. While the $\ell=1$ orbitals would be most susceptible to confinement effects, in $^{35}$S they appear reasonably-well bound. Similarly, though the $0f_{5/2}$ neutron orbital centroid is likely above the $S_n$, it would still be confined by a few MeV.

\section{Summary and Conclusions}\label{sec:sum}

The single-neutron overlaps, strengths, and centroid energies were determined from the cross sections measured in the $^{34}$S($d$,$p$) reaction at 8~MeV/$u$ for the $0f_{7/2}$, $1p_{3/2}$, $1p_{1/2}$ and $0f_{5/2}$ neutron orbitals. The experiment took place at the Florida State University John D. Fox Laboratory and utilized the Super-Enge Split-Pole Spectrograph to analyze the outgoing reaction protons. In a complementary measurement, the CeBrA demonstrator~\cite{ref:Conley2024} provided $\gamma$-ray detection around the target location for the same reaction. In total, over 20 previously established or candidate states in $^{35}$S were observed up through $E_x \approx 7.7$~MeV.

Primarily focused on the extracted C$^2$S below $E_x < 5.1$~MeV, a consistent picture of the vacancies for the $0f_{7/2}$, $1p_{3/2}$ and $1p_{1/2}$ orbitals emerged. Each showed fragmentation of the their strength across two or more individual states. The single-neutron energy centroids showed a reduced energy spacing at $N=28$, consistent with the situation in the more neutron-rich $Z=16$ isotopes. The neutron $1p_{3/2}$-$1p_{1/2}$ energy spacing, established with the proper $1p_{1/2}$ strengths, defined the spin-orbit energy difference at $Z=16$, $N=19$. A lower limit on the $0f_{5/2}$ centroid energy placed it above the $1p_{1/2}$ orbital. This was in agreement with the energy ordering in other nearby isotopes including those of the neutron-rich Ca isotopes.

\bmhead{Acknowledgements}

This work was supported by the National Science Foundation (NSF), United States, under Grant No. PHY-2012522 (FSU) and is based upon work supported by the U.S. Department of Energy, Office of Science, Office of Nuclear Physics, under Contract No. DE-AC02-06CH11357 (Argonne). Targets were provided by the Center for Accelerator Target Science (CATS) at Argonne National Laboratory.

\bibliography{s34dpbib}

\end{document}